\documentclass{article}
\usepackage[activate=true,
    final,
    tracking=true,
    kerning=true,
    spacing=true,
    factor=1100,
    stretch=10,
    shrink=10]{microtype}
    
\usepackage{amssymb}
\usepackage{bbding}
\usepackage{pifont}
\usepackage{utfsym}
\usepackage{fontawesome}
\usepackage{dcase2021_techrep,amsmath,graphicx,url,times,booktabs, tabularx, siunitx}

\microtypecontext{spacing=nonfrench}

\title{Surrey System For DCASE 2022 Task 5: Few-shot Bioacoustic Event Detection With Segment-Level Metric Learning}

\name{
      Haohe Liu$^{1}$,
      Xubo Liu$^{1}$,
      Xinhao Mei$^{1}$, 
      Qiuqiang Kong$^{2}$,
      Wenwu Wang$^{1}$,
      Mark D. Plumbley$^{1}$
}
\address{$^1$ Centre for Vision, Speech, and Signal Processing~(CVSSP), University of Surrey, UK\\ 
        $^2$ Speech, Audio, and Music Intelligence (SAMI) Group, ByteDance, China \\
}

\begin{document}

\ninept
\maketitle

\begin{sloppy}

\begin{abstract}
Few-shot audio event detection is a task that detects the occurrence time of a novel sound class given a few examples. In this work, we propose a system based on segment-level metric learning for DCASE 2022 challenge few-shot bioacoustic event detection~(task 5).
We make better utilization of the negative data within each sound class to build the loss function, and use transductive inference to gain better adaptation on the evaluation set. For the input feature, we find the per-channel energy normalization concatenated with delta mel-frequency cepstral coefficients to be the most effective combination. We also introduce new data augmentation and post-processing procedures for this task. Our final system achieves an f-measure of 68.74 on the DCASE task 5 validation set, outperforming the baseline performance of 29.5 by a large margin. Our system is fully open-sourced\footnote{\url{https://github.com/haoheliu/DCASE_2022_Task_5}}.
\end{abstract}

\begin{keywords}
few-shot learning, transductive inference, metric learning, audio event detection
\end{keywords}

\section{Introduction}
\label{sec:intro}
Few-shot learning~(FSL)~\cite{snell2017prototypical} is a machine learning problem that aims to generalize the system to novel classes after supervised learning with the data from known classes. Sound event detection~\cite{mesaros2021sound} is a task that locates the onset and offset of certain sound classes. By combining the idea of FSL with sound event detection~\cite{wang2020few}, the system can detect the new sound class with only a few labeled samples. This technique is useful for audio data labeling, especially when the user needs to detect a new type of sound.

In the previous DCASE 2021 task 5 challenge, the submitted systems~\cite{yang2021few, tang2021two, zhang2021few, anderson2021bioacoustic} have achieved promising results on the evaluation set. Most of the systems use a prototypical network~\cite{snell2017prototypical} as the main architecture. Yang et al.~\cite{yang2021few} propose a mutual learning framework to improve the feature extractor. Tang et al.~\cite{tang2021two} propose to use embedding propagation to learn a smoother manifold in the latent space. Data augmentations such as spec-augment and mixup are used in the system described in~\cite{zhang2021few, anderson2021bioacoustic}.

In previous works, the negative segment in each audio file is not fully utilized. Here negative segment means the audio segments that do not contain the target sound event. Most metric-learning-based studies~\cite{yang2021few, tang2021two} optimize the model only using the labeled positive data, by grouping and separating the latent prototypes of the same and different positive classes, respectively. However, the positive segments also need to be distinguishable from the negative segments within the same audio file. We propose to use negative segments to enhance the metric learning to address this problem.

This report describes the prototypical-network based system we submitted to the DCASE challenge 2022 task 5~\cite{morfi2021few,liu2022segment}. Our system builds upon a prototypical network and proposes new methods for the input features, metric learning, evaluation data preprocessing, and post-processing. 


To increase the generalization ability of our system, we merge the training set with the animal sound in AudioSet~\cite{gemmeke2017audio}. 
We conduct studies on a different combinations of audio features to choose the best feature for this task. Features includes log-mel spectrogram~(MEL), per-channel energy normalization~(PCEN), mel-frequency cepstral coefficients~(MFCC), and delta-MFCC~($\Delta$MFCC). We found using PCEN and delta-MFCC together yields the best average \text{f-measure} score on the validation set.
To make full use of the labeled few-shot data, we integrate transductive inference into our metric learning framework. We treat each evaluation file as a sound class and jointly optimize it with the training dataset. 
Our proposed system achieves a \text{f-measure} of 68.74 on the DCASE task 5 validation set. 

This technical report will be organized as follows. Section~\ref{sec:overview} provides an overview of our system. Section~\ref{sec:methodology} introduces the methods we proposed to improve our system. Section~\ref{sec:experiments} discussed the experimental setup and the four systems we submitted to the challenge. Section~\ref{sec:result} reports the result of the validation set and the ablation studies. Section~\ref{sec:conclusions} summarize this work and provide a conclusion.


\section{System Overview}
\label{sec:overview}

We build our system with a prototypical network, which is widely used for metric-based few-shot learning. 
The objective of our system is properly mapping different sounds into a latent embedding within a high-dimensional space, where similar sounds have a closer distance. 

We use episodic training~\cite{li2019episodic} to optimize our system in an N-way-K-shot way, in which we set $N=10$ and $K=5$. This means each training batch will select data from N classes. And for each class, the system will select $2\times K$ segments from the dataset, and treat $K$ segments as support segments and other $K$ segments as query segments. Then both segments are processed by a feature extraction network, which outputs latent embeddings. By averaging the embedding of $K$ support and $K$ query segments for class $i$, we obtain a query prototype $x^{q}_{i}$ and a support prototypes $x^{s}_{i}$. The system is optimized by minimizing the distance between query prototypes and support prototypes for each class. 
To build a robust latent space and to generalize better to the new class, we introduce the metric learning with negative samples and the transductive inference approach in Section~\ref{sec:negative-metric-learning} and~\ref{sec:transductive-inference}. 

During evaluations, only fivethe system uses $K$ labeled positive events to build positive and negative prototypes, which are the latent representation of positive and negative segments in one audio file. Here a positive sound event can contain multiple positive segments. If the negative segment in the labeled part is too short, we will use a spectrum-similarity-based algorithm~(Section~\ref{sec:negative-prototypes}) to identify the possible negative segment in the remaining unlabelled audio. To classify the unlabelled part, the system will first segment it with an adaptive segment length and build a query set. Then the query set can be classified by calculating the distance with prototypes. We introduce the way we build the positive and negative prototypes in Section~\ref{sec:negative-prototypes}. 

Post-processing is important to the performance of the system. We introduce a splitting-merging-filtering-based post-processing method designed to fully utilize the information provided in the first $K$ labeled events, such as minimum negative segment length. We also use various data augmentation techniques and external training data from AudioSet during model training, which will be discussed in section~\ref{sec:experiments}.


\section{Methodology}
\label{sec:methodology}
\subsection{Feature extraction network}
Our feature extraction network $f_{\theta}$ is a convolutional neural network~(CNN) based architecture that maps the audio feature $s$ into a latent embedding $x$:

\begin{equation}
    f_{\theta}: s \rightarrow x
\end{equation}

Similar to the architecture proposed by~\cite{kong2020panns}, the network $f_{\theta}$ consists of three convolutional blocks with hidden channels of 64, 128, and 64. Each convolutional block consists of three two-dimensional CNN layers with batch normalization and leaky rectified linear unit activations~\cite{xu2015empirical}. As a common trick in CNN-based network~\cite{liu2020channel, liu2021voicefixer}, we apply $2\times 2$ max-pooling after each block. The input and output of each convolutional block have a residual connection processed by a downsampling CNN layer. In order to maintain the same output dimension with different input lengths, we apply an adaptive average pooling at the end of the network. The final output feature map after adaptive pooling is a $C\times T\times F$ size block, which is also the final latent embedding. 

\subsection{Segment-level metric learning}
\label{sec:negative-metric-learning}

In a similar way to ~\cite{wang2020few}, we divide the audio features into segments with equal lengths for metric learning. Each segment has a unique label indicating its sound class and whether it is a positive or negative event.
As illustrated in the left part of Figure~1, previous methods typically optimize the model only using the labeled positive classes, by minimizing the distance between the query and support prototypes that have the same label. However, these systems only use the labeled positive classes, which are limited in duration. 
By comparison, negative samples usually contain a wider variety of sounds and are easier to collect.
In addition, during inference, the system will need to discriminate between positive and negative samples. This objective is not directly optimized during training with only positive data. 

\begin{figure}[htbp]
    \centering
    \label{fig:negative-enhanced-contrastive-learning}
    \includegraphics[page=1, width=\columnwidth]{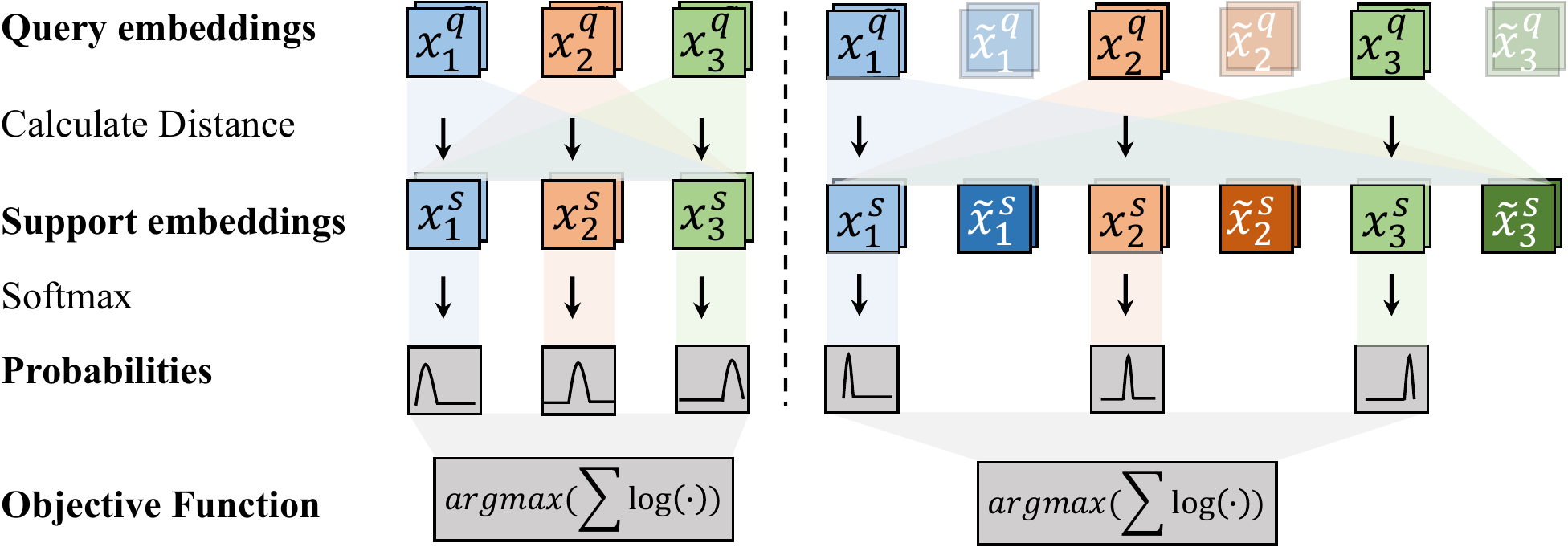}
    \caption{The left side is the baseline method that performs metric learning in a 3-way-2-shot way. The baseline method only uses positive classes. The right side illustrates the improved metric learning with the negative samples method we used in our system. $x_{i}$ and $\widetilde{x}_{i}$ stand for the positive and negative samples of class $i$. Transparent color means the samples are not actually used.}
    \vspace{-5mm}
\end{figure}

So, to better discriminate between negative and positive segments within the audio, we propose to perform a segment-level metric learning approach with the negative segments. 
In on the probabilities of the positive classes, the query will also need to calculate the distance with the negative support embeddings. The comparison between positive and negative segments will provide the model with more contrastive information on building the latent space. Note that we do not optimize the probabilities of the negative query prototypes, because negative samples are not guaranteed to have the same sound for classification. 



\subsection{Transductive inference}
\label{sec:transductive-inference}
We adopt a transductive inference~\cite{boudiaf2021few} approach during training. Specifically, our model will be optimized both on the development set and the partially labeled evaluation set. However, the labeled segment in the evaluation set is much shorter than in the training set. To balance between these two sets, we uniformly sample $N$ sound class from both sets during training.

\subsection{Loss function}
We optimize $f_{\theta}$ by maximizing the probability that $x^{q}_{i}$ belongs to class $i$. Specifically, we first calculate the probability distribution of $x^{q}_{i}$ among $N$ possible classes by Equation~\ref{eq:distance}. Then we maximize the probability that $x^{q}_{i}$ belongs to class $i$ by summing and maximizing the diagonal elements of $\mathbf{D}$, given by

\begin{equation}
    \label{eq:distance}
    \mathbf{D}_{i,j} = \sqrt{\Sigma((x^{q}_{i}-x^{s}_{j})^{2})},~\mathbf{\widetilde{D}}_{i,:} = \textrm{log}(\textrm{Softmax}(\mathbf{D}_{i,:})),
\end{equation}

\begin{equation}
    l = \textrm{argmax}_{\theta}(\Sigma(\mathbb{I} \odot \mathbf{\widetilde{D}})),
\end{equation}
where $l$ is the loss function, $\mathbb{I}$ is the identity matrix, $\odot$ stands for element-wise multiplication, and $\Sigma$ means summing all the elements in the matrix.

\subsection{Adaptive segment length}
\label{sec:ada-seg-len}
During the inference process, the audio file will be divided into segments with adaptive segment lengths. The labeled positive segments will be used to build positive prototypes. The segments after the last labeled positive event, which are unlabeled segments, will be treated as the query set. Both the query set and labeled negative segments can be used to build the negative prototypes. 

Sounds of different animal or bird species have drastically different lengths of vocalization, ranging from 30 milliseconds to 5 seconds. To have robust detection result, we set different window lengths and hop lengths for each audio file based on $t_{\max}=\max(t_{1},...,t_{K})$, as shown in Table~\ref{tab:seg-len-setting}, where $t_{1},...,t_{K}$ denotes the duration of the $K$ labelled positive events. We set the hop length as one-third of the window length. 

\begin{table}[tbp]
\label{tab:seg-len-setting}
\centering
\begin{tabular}{c|c|c|c|c|c}
\hline
$t_{\max}$ (s)  & {[}0,0.1{]} & (0.1,0.4{]} & (0.4,0.8{]} & (0.8,3.0{]} & (3.0,+inf) \\ \hline
Length &    8    &  $t_{\max}$       &  $t_{\max}$ / 2         & $t_{\max}$ / 4          & $t_{\max}$ / 8        \\ \hline
\end{tabular}
\caption{The window length we use during segmenting the evaluation and validation audio file for different values of $t_{\max}$.}
\end{table}

\subsection{Positive and negative prototypes}
\label{sec:negative-prototypes}
We calculate the positive prototype by simply averaging the embeddings of the labeled positive segments because we assume positive events do not contain too much variety. By comparison, building negative prototypes is more tricky for the following challenges:

\begin{enumerate}
    \item Negative segments can contain many different kinds of sounds. So simply averaging all the negative embeddings will result in a sub-optimal representation of negative prototypes.
    \item The duration of labeled negative data is limited in some test files. In this case, some negative events appear later in the query set may not be included in the negative prototype.
\end{enumerate}

Note that the baseline system uses the average embedding of the entire audio as the negative prototype, under the assumption that the positive event is sparse. However, this assumption is not true in most of the evaluation files. Building a negative prototype in this way can lead to a degraded result.

To address the first challenges, we choose to run our evaluation six times, each with different 30 randomly selected negative segment embeddings. And we mean the predicted probabilities across time as the final predictions. In this case, the negative prototype in each run can have a chance to represent different sound classes.

For the second challenge, we propose a negative sample searching algorithm~(Figure~\ref{fig:negative-sample-searching}) to find more negative segments to build robust negative prototypes. The algorithm includes a frequency bins weighting step and a frequency pattern matching step.

\begin{figure}[tbp]
    \centering
    \label{fig:negative-sample-searching}
    \includegraphics[page=2, width=0.9\columnwidth]{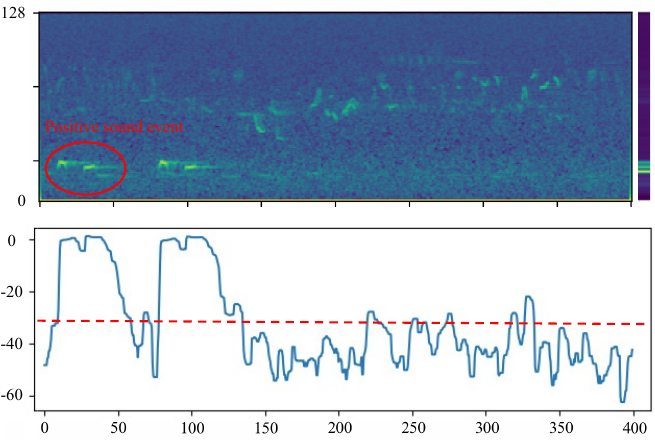}
    \caption{The visualization of the negative sample searching algorithm. In the first row, the left picture is the log-mel spectrogram of an audio clip, and the picture on the right side is the mel-frequency bins weight. The second row shows the frequency pattern matching score. The segment with a score under the red dashed line will be used as negative segment.}
    \vspace{-5mm}
\end{figure}

The frequency bins weighing operation aims to find the frequency band that is most likely to contain the target sound event so that we can find the negative event more accurately. This operation is based on two assumptions. First, we assume that bioacoustic sound events for a certain class~(e.g., owl vocalization) do not have too many variations in frequency. Localizing the most important frequency band can therefore assist the model to ignore interference sounds in other frequency bands. Second, since onset and offset time mark the start and end of a sound event, we assume that the frequency bins that have more energy fluctuations before and after one positive event are the frequency bands that contain target sound events. We calculate frequency bin weight $w$ by calculating the energy difference between the positive segments and the negative segments before and after these events, and then $w$ is normalized between zero and one.

Then we perform frequency pattern matching to locate possible negative samples. We first multiply the linear mel-spectrogram with the frequency bin weights. Then we calculate the average of positive time steps as the frequency pattern template. Finally, we calculate the scale-invariance signal-to-noise ratio~(SISNR)~\cite{le2019sdr} between the template and each time step in the weighted mel-spectrogram. If the result of a time step is smaller than a threshold, we will treat it as a negative event. The threshold is calculated using the minimum SISNR value between the template and positive segments. Here we use the scale-invariance version of SNR to overcome the situation where the energy of a positive event changes.

\subsection{Post processing}

We determine the post processing strategy of a sound class based on maximum length of positive event $t_{\max}$ and minimum length of negative event $\widetilde{t}_{\min} = \min(\widetilde{t}_{1},...,\widetilde{t}_{K})$, where $\widetilde{t}_{1},...,\widetilde{t}_{K}$ is the durations of the $K$ labelled negative segments. We perform the following steps for post processing. Note that most parameters in the following steps are chosen by experience.

\begin{table}[htbp]
\centering
\label{tab:seglen-negative}
\begin{tabular}{c|c|c|c}
\hline
$\widetilde{t}_{\min}~(s)$      & {[}0,~8{]} & (8,~100{]} & (100,~+inf) \\ \hline
Length & 8         & $\widetilde{t}_{\min} / 2$         & 100        \\ \hline
\end{tabular}
\caption{The window length we use during short negative segment detection for different values of $\widetilde{t}_{\min}$.}
\end{table}

\begin{table*}[t!]
\label{tab:final-system}
\begin{tabular}{c|ccc|ccc|ccc|ccc}
\hline
 &
  \multicolumn{3}{c|}{HB} &
  \multicolumn{3}{c|}{ME} &
  \multicolumn{3}{c|}{PB} &
  \multicolumn{3}{c}{Overall} \\ \hline
 System &
  \multicolumn{1}{c|}{Pre (\%)} &
  \multicolumn{1}{c|}{Rec (\%)} &
  F (\%) &
  \multicolumn{1}{c|}{Pre (\%)} &
  \multicolumn{1}{c|}{Rec (\%)} &
  F (\%) &
  \multicolumn{1}{c|}{Pre (\%)} &
  \multicolumn{1}{c|}{Rec (\%)} &
  F (\%) &
  \multicolumn{1}{c|}{Pre (\%)} &
  \multicolumn{1}{c|}{Rec (\%)} &
  F (\%) \\ \hline
Liu\_S1 &
  \multicolumn{1}{c|}{\textbf{98.87}} &
  \multicolumn{1}{c|}{79.31} &
  \textbf{88.01} &
  \multicolumn{1}{c|}{78.95} &
  \multicolumn{1}{c|}{\textbf{86.54}} &
  82.57 &
  \multicolumn{1}{c|}{48.18} &
  \multicolumn{1}{c|}{28.70} &
  35.97 &
  \multicolumn{1}{c|}{68.90} &
  \multicolumn{1}{c|}{50.84} &
  58.51 \\ \hline
Liu\_S2 &
  \multicolumn{1}{c|}{87.03} &
  \multicolumn{1}{c|}{\textbf{83.08}} &
  85.01 &
  \multicolumn{1}{c|}{65.00} &
  \multicolumn{1}{c|}{50.00} &
  56.52 &
  \multicolumn{1}{c|}{36.13} &
  \multicolumn{1}{c|}{30.00} &
  32.78 &
  \multicolumn{1}{c|}{54.99} &
  \multicolumn{1}{c|}{45.89} &
  50.03 \\ \hline
Liu\_S3 &
  \multicolumn{1}{c|}{94.50} &
  \multicolumn{1}{c|}{59.67} &
  73.15 &
  \multicolumn{1}{c|}{65.85} &
  \multicolumn{1}{c|}{51.92} &
  58.06 &
  \multicolumn{1}{c|}{35.40} &
  \multicolumn{1}{c|}{17.39} &
  23.32 &
  \multicolumn{1}{c|}{55.54} &
  \multicolumn{1}{c|}{32.08} &
  40.67 \\ \hline
Liu\_S4 &
  \multicolumn{1}{c|}{97.95} &
  \multicolumn{1}{c|}{79.46} &
  87.74 &
  \multicolumn{1}{c|}{86.27} &
  \multicolumn{1}{c|}{84.62} &
  85.44 &
  \multicolumn{1}{c|}{57.52} &
  \multicolumn{1}{c|}{27.66} &
  37.36 &
  \multicolumn{1}{c|}{\textbf{76.56}} &
  \multicolumn{1}{c|}{49.54} &
  60.16 \\ \hline
Liu\_S0 &
  \multicolumn{1}{c|}{82.20} &
  \multicolumn{1}{c|}{82.33} &
  82.26 &
  \multicolumn{1}{c|}{\textbf{95.56}} &
  \multicolumn{1}{c|}{82.69} &
  \textbf{88.66} &
  \multicolumn{1}{c|}{\textbf{59.88}} &
  \multicolumn{1}{c|}{\textbf{42.17}} &
  \textbf{49.49} &
  \multicolumn{1}{c|}{76.28} &
  \multicolumn{1}{c|}{\textbf{62.56}} &
  \textbf{68.74} \\ \hline
\end{tabular}
\caption{The precision, recall, and f-measure of each subset in the validation set. S1,S2,S3,S4 are four systems we finally submitted to the challenge. We do not submit S0 on considering that the model might overfit to the validation dataset.}
\end{table*}

After detecting the onset and offset of short negative segments, we perform splitting on the positive segment. This operation is only performed when the length of the detected positive event is longer than $2t_{\max}$ and a negative segment is detected within the positive event.


We determine the threshold segment length of a sound class with the duration of the few labeled positive and negative segments $\mathbf{t}$ and $\mathbf{t^{\prime}}$. During filtering, the minimal segment length is $0.4 * \bar{\mathbf{t}}$. The merging operation will be performed if two segments have a total duration less than $0.8 * \bar{\mathbf{t}}$ and the negative segment between them is shorter than the minimal duration in $\mathbf{t^{\prime}}$. The system will consider to perform splitting operation if the segment length is longer than $2.0 * \bar{\mathbf{t}}$. And we design the negative segment mining approach to determine where to perform the splitting.


The essential reason why the model cannot locate the negative segment between two adjacent positive sound events is because of the length of the query window. There is a trade-off during inference. If the window length is not small enough, the query sample could contain positive and negative samples at the same time, and the negative part will not be effectively detected. However, if the window length is too small, the model tends to have too many false positives that will affect the final score. 

To address this problem, we propose negative segment mining operations, which means detecting the possible negative segment. Since the false positive is not important in this case, we can use a small query window length that is capable of detecting small intervals between two positive sound events. The detected negative intervals will be used for splitting long positive segments if the segment length is longer than $2.0 * \bar{\mathbf{t}}$.

\section{Experiments}
\label{sec:experiments}

\subsection{Dataset}

\textbf{Challenge official dataset} DCASE 2022 task 5 dataset contains a training set, a validation set, and an evaluation set. The training and validation set are both fully labeled. The evaluation set is provided only with the labels of the first five positive events.

\noindent
\textbf{AudioSet}
\label{sec:AudioSet-external_data} AudioSet~\cite{gemmeke2017audio} is a large-scale dataset for audio research~\cite{kong2020panns, kong2021speech}. The training set of the DCASE 2022 task 5 challenge only contains 47 different sound classes. To enhance the latent space with a wider variety of sounds, we choose to build an extra training set using the strongly labeled part of the AudioSet dataset. To avoid the domain mismatch problem, we only utilize the sound labels that are related to animal vocalizations. We only use animal sounds that do not overlap with other non-animal sounds. We will not use the data if the total duration of non-overlapping parts is less than two seconds within the audio because we consider in this case the sound event is not enough to use. 





\subsection{Evaluation metric}
We use the f-measure score, the official evaluation metric\footnote{https://github.com/c4dm/dcase-few-shot-bioacoustic} provided by the organizers of DCASE task 5, as our main evaluation metric. We also report the accuracy and recall.

\subsection{Experimental setup}
Following the prior works~\cite{yang2021few}, all the audio data are resampled into 22.5 kHz sampling rate. The input feature of our system is the stack of PCEN~\cite{wang2017trainable} and delta-MFCC~\cite{5709752} features. In the short-time Fourier transform, we set the window length as 1024 and hop size as 256. We set the mel dimension as 128. The input length of our model during training is 0.2 seconds, and the output embedding dimension is 2048. All the experiments use an initial learning rate of 0.001 with 0.65 exponential decay every 10 epochs. We will stop model training if the validation accuracy does not improve for 10 consecutive runs. And the model with the best validation accuracy is used for evaluation. For model evaluation, we consider the negative segments with a total duration of fewer than 2.0 seconds are not enough to build the negative prototype. And in this case, we will use the negative sample searching algorithm mentioned in Section~\ref{sec:negative-prototypes}.

Different from the baseline that preprocesses and segments the training data with a certain hop length, we implement a dynamic approach that generates training data on the fly. During training, we will randomly select the starting point of the training segment. In this case, the training data can be fully utilized.

\section{Result}
\label{sec:result}

\begin{table}[htbp]
\label{tab:comparison_with_baseline}
\begin{tabular}{c|c|c|c}
\hline
\multicolumn{1}{c|}{Method} & \multicolumn{1}{c|}{Precision (\%)} & \multicolumn{1}{c|}{Recall (\%)} & \multicolumn{1}{c}{F-measure (\%)} \\ \hline
Template Matching  & 2.42  & 18.32   & 4.28  \\ \hline
Prototypical~\cite{morfi2021few} & 36.34 & 24.96  & 29.59 \\ \hline
\textbf{Proposed}         & 76.28    & 62.56     & 68.74 \\ \hline
\end{tabular}
\caption{Comparison with baseline template matching and prototypical network methods}
\end{table}

The performance of our system on the validation set is reported in Table~\ref{tab:comparison_with_baseline}. The F-measure score of the prototypical network baseline is 29.59. Our system outperforms the baseline by a large margin with an F-measure score of 68.74, using a threshold of 0.95.
We submit four systems for the DCASE task 5 challenge. The performance of each system on each validation subset is reported in Table~\ref{tab:final-system}. We select systems with different performances to avoid cherry-picking models that only work well on the validation set. All four systems use the same configurations except the settings shown in Table~\ref{tab:setting}. We tried different post-processing methods and training data in these four systems.

\begin{table}[htbp]
\label{tab:setting}
\begin{tabular}{c|c|c|c|c|c|c}
\hline
    System    & Pad & Split & Merge & AudioSet & Threshold      & Ens \\ \hline
Liu\_S1 & \ding{55}   & \checkmark     & \ding{55}     & \ding{55}        & 0.5 - 0.995 & \checkmark        \\ \hline
Liu\_S2 & \checkmark   & \ding{55}     & \ding{55}     & \checkmark        & 0.6         & \ding{55}        \\ \hline
Liu\_S3 & \checkmark   & \checkmark     & \checkmark     & \ding{55}        & 0.95        & \ding{55}        \\\hline
Liu\_S4 & \ding{55}   & \ding{55}     & \ding{55}     & \ding{55}        & 0.6         & \ding{55}  \\   \hline

\end{tabular}
\caption{The setting of each submitted system. Ens stand for performing ensamble on different thresholds (0.5,0.7,0.9,0.99,0.995).}
\vspace{-5mm}
\end{table}

\section{Conclusions}
\label{sec:conclusions}
This technical report describes the system we submitted to the DCASE 2022 challenge task 5. Our system use segment-level metric learning with negative segments, negative segment searching, and post-processing. The experimental result indicates our system can improve the baseline prototypical network by a large margin.



\section{ACKNOWLEDGMENT}
\label{sec:ack}
This research was partly supported by a PhD scholarship from the Centre for Vision, Speech and Signal Processing (CVSSP), Faculty of Engineering and Physical Science (FEPS), University of Surrey and BBC Research and Development, and a Research Scholarship from the China Scholarship Council (CSC) No. 202006470010. For the purpose of open access, the authors have applied a Creative Commons Attribution (CC BY) licence to any Author Accepted Manuscript version arising.


\bibliographystyle{IEEEtran}
\bibliography{refs}

\end{sloppy}
\end{document}